\documentclass{article} 
\usepackage[final]{colm2026_conference}

\usepackage{microtype}
\usepackage{hyperref}
\usepackage{url}
\usepackage{booktabs}
\usepackage{graphicx}
\usepackage{enumitem}

\usepackage{lineno}

\definecolor{darkblue}{rgb}{0, 0, 0.5}
\hypersetup{colorlinks=true, citecolor=darkblue, linkcolor=darkblue, urlcolor=darkblue}

\title{Safety Benchmarks as Measurement Instruments: \\ A Crossed Evaluation of LLM Harm Benchmarks}

\author{Emaan Bilal Khan, Dylan Hadfield-Menell \\
MIT CSAIL\\
}

\begin{document}

\ifcolmsubmission
\linenumbers
\fi
\raggedbottom

\maketitle

\begin{abstract}
Safety benchmarks increasingly shape how language models are compared, released, audited, and governed. These scores are often treated as model properties, even when the benchmarks producing them rely on different prompt distributions, judge models and rubrics. We test how much these operational choices matter by studying 5 popular safety benchmarks that all draw from major provider usage policies. Using a fully crossed 5 × 5 design across 13 open-source models, we recombine each benchmark’s prompt set with each scorer to separate prompt-side and scorer-side effects. We find that disagreement is structured by measurement protocol: scorer-side protocols introduce more variation than prompt sets, benchmark choice substantially shifts rankings for middle-ranked and domain fine-tuned models, and cross-benchmark validity varies by harm category. In an exploratory multitrait-multimethod analysis, benchmarks rank models similarly for concrete harmful-use categories such as fraud or adult content, but not for contextual categories such as privacy, political campaigning, and hate. These findings illustrate that safety benchmark scores are protocol-dependent measurements rather than intrinsic model properties, motivating reporting, calibration, and governance practices that tie safety claims more tightly to the operational choices that produce them.
\end{abstract}

\section{Introduction}

As generative AI use cases expand, benchmarking tools have become a primary source of empirical evidence for model capability, deployment readiness, and safety. Benchmark results are routinely cited in provider system cards to demonstrate robustness to harmful requests \cite{OpenAI_GPT5_System_Card_2025, Anthropic_Claude4_System_Card_2025}, used by independent auditors to inform deployment recommendations, and increasingly incorporated into regulatory frameworks such as the EU AI Act and the NIST AI Risk Management Framework as part of risk assessment processes for general-purpose AI systems \cite{EUAIAct2024, NIST_GAI_Profile_2024}. To meet these growing demands, the machine learning community has developed a diverse ecosystem of safety benchmarks, each designed to measure different aspects of large language model (LLM) behavior, including harmfulness, refusal behavior, jailbreak robustness, and policy compliance \cite{ni2025surveylargelanguagemodel}.

Unlike accuracy on a closed-form task, however, harmfulness is not a directly observable property of a model output. In many cases, there is no objective ground-truth label indicating whether a response is ``unsafe.'' Instead, harmfulness must be inferred through observable behavioral indicators, such as compliance with a malicious request, the provision of actionable details, or explicit policy violations. At scale, these judgments are typically operationalized through automated evaluation pipelines, including LLM-as-a-judge protocols \cite{gu2025surveyllmasajudge}. A growing body of work documents substantial reliability concerns with these methods \cite{schroeder2025trustllmjudgmentsreliability}. Seemingly minor operational choices, such as response ordering, evaluator prompts, judge model, even stylistic features of the response itself, can substantially alter reported safety outcomes \cite{chen2025saferluckierllmssafety, zhu2025advprefixobjectivenuancedllm}. Benchmark scores may therefore reflect not only differences in underlying model behavior, but also artifacts of the evaluation procedure used to produce them.


These reliability failures validate recent measurement-theoretic critiques of generative AI evaluation, most prominently Wallach et al.'s framework distinguishing the background concept, systematized concept, measurement instrument, and reported score as separate levels at which validity must be interrogated \cite{wallach2025position, salaudeen2025}. While such position papers outline vital structural frameworks, the field lacks empirical contributions that isolate and measure these vulnerabilities across evaluation pipelines in current use. Without a decomposition of how prompt sets, harm categories, and scoring protocols interact, the question of when benchmark scores are comparable and what claims they can validly support remains open.

In this paper, we present a systematic interrogation of measurement choices within a high-stakes family of safety benchmarks. We examine five prominent benchmarks (HarmEval, HEx-PHI, JailbreakBench, StrongREJECT, and ForbiddenQuestions), across 13 open-source large language models spanning instruction-tuned, abliterated, and fine-tuned training regimes \cite{banerjee2024harmeval, qi2024hexphi, chao2024jailbreakbench, souly2024strongrejectjailbreaks, shen2024fq}. While each benchmark draws their harm categories from major provider usage policies, they operationalize that shared anchor through different prompt sets, judge models, rubrics, and scoring thresholds. Rather than evaluating each benchmark only in its native configuration, we conduct a fully crossed $5 \times 5$ evaluation where each benchmark contributes a prompt set and a scorer, and every prompt set is evaluated under every scorer. This allows us to separate prompt-side from scorer-side variation and examine model-rank sensitivity and category-level agreement across benchmarks. 

Our findings reveal several important empirical patterns: (1) We show that scorer-side protocols introduce more variation in safety estimates than the underlying prompt sets in our sample, showing that the choice of judge model, rubric, and score scale are central components of the measurement instrument rather than supplementary implementation details. (2) While model safety rankings remain stable at the extremes of the distribution, we highlight that fine-tuned models exhibit substantial rank instability across evaluation setups. (3) A preliminary Multi-Trait Multi-Method analysis shows that construct validity varies sharply by harm category: concrete harms such as fraud or adult content show stronger cross-benchmark validity, while privacy, political campaigning, and hate/harassment/violence do not support straightforward cross-benchmark comparison.

These results offer concrete empirical guidance supporting more robust evaluation science efforts. While prior conceptual work has highlighted the need for well-defined constructs, greater methodological transparency and validity evidence, our crossed prompt-set $\times$ scorer design enables us to demonstrate how these vulnerabilities manifest in practice. 
Leveraging these outcomes, we examine their practical implications, outlining how safety benchmarks should be reported, designed, and maintained over time for downstream users to know what claims their scores can and cannot support.

\section{Related Work}

\textbf{Measurement validity in AI evaluation.} A growing literature applies measurement theory from the social sciences to AI benchmarks. Jacobs and Wallach introduce measurement modeling as a framework for fairness evaluation, arguing that many fairness harms stem from mismatches between constructs and their operationalizations \cite{Jacobs_2021}. Blodgett et al.\ apply this lens to four widely used fairness benchmarks, cataloguing pitfalls where benchmarks lack clear articulations of what is being measured \cite{blodgett-etal-2021-stereotyping} . More recent work extends these frameworks \cite{raji2021aiwideworldbenchmark}, proposing a four-level framework that separates background concept, systematized concept, measurement instrument, and measurement as distinct levels at which validity must be interrogated \cite{wallach2025position}. Salaudeen et al. similarly argue that validity depends on the claim a score is used to support, with broader claims requiring stronger evidence \cite{salaudeen2025}. Bean et al. audit 445 capability benchmarks against this framework and find nearly all have weaknesses in concept articulation, operationalization, or claim support \cite{bean2025measuringmattersconstructvalidity}. Our work operationalizes these frameworks for a specific high-stakes benchmark family, empirically decomposing how design choices in widely interchanged safety benchmarks affect the validity of the claims those benchmarks are used to support.

\textbf{Reliability concerns in safety benchmarking.} A parallel line of work documents specific reliability failures in LLM safety benchmarks and the LLM-as-a-judge protocols that underlie them. Ren et al.\ show that scores on many safety benchmarks correlate strongly with general capability, raising the possibility that ``safety'' as currently measured is not empirically separable from broader model quality \cite{ren2024safetywashingaisafetybenchmarks}.  \cite{subramonian-etal-2023-takes} report that NLP practitioners disagree about what benchmarks measure even within a single task family. Within safety evaluation specifically, the choice of automated judge has been shown to materially shift reported outcomes: judges overestimate attack success rates by factors of three to four depending on the target model \cite{zhu2025advprefixobjectivenuancedllm}, an apologetic prefix alone can shift judge preferences by up to 98\% \cite{chen2025saferluckierllmssafety}, and refusal rates for the same model vary from 13\% to 58\% across over-refusal benchmarks \cite{zhang2025falserejectresourceimprovingcontextual}. Concurrent work using Generalizability Theory reports near-zero cross-benchmark generalizability, with model and benchmark interaction accounting for the majority of variance \cite{gringras2026safetyscaffoldingevaluationconditions}. These results consistently point to substantial disagreement among safety benchmarks, but treat individual benchmarks as black-box instruments without decomposing the contribution of their components. Our crossed design isolates prompt-side and scorer-side effects at the protocol level, and our MTMM analysis breaks down category-level agreement that prior work has not explored. 

\section{Methodology}

\textbf{Benchmark protocols.} We evaluate five LLM safety benchmarks: HarmEval, HEx-PHI, JailbreakBench, StrongREJECT, and ForbiddenQuestions \cite{banerjee2024harmeval, qi2024hexphi, chao2024jailbreakbench, souly2024strongrejectjailbreaks, shen2024fq}. We selected these benchmarks because they all derive their harm categories from major provider usage policies, and so appear to share the same definitional source for the constructs they assess. Each benchmark nonetheless operationalizes unsafe model responses differently. We treat each as a measurement instrument with two separable components: a prompt set, which samples behaviors from a harm taxonomy, and a scorer-side protocol, which maps a model response to a safety judgment. The scorer-side protocol bundles the judge model, scoring prompt, rubric, score scale, and any in-context examples. Full factorial ablations separating the effect of these subcomponents are left for future work.

The five benchmarks differ substantially in their scorer design (Table \ref{tab:scorer_design}), spanning binary judges with no policy reference (HarmEval), policy-anchored Likert scales (HEx-PHI), strict specificity thresholds (JailbreakBench), composite refusal-specificity-convincingness scoring (StrongREJECT), and few-shot answer-detection (ForbiddenQuestions). All five use off-the-shelf judge models (GPT-4o-mini, Llama-3-70B-Instruct, and ChatGLM2-6B; Table 1). 

\begin{table}[t]
\vspace{-0.4em}

\scriptsize
\setlength{\tabcolsep}{3pt}
\renewcommand{\arraystretch}{0.9}

\begin{tabular}{p{2.2cm} p{3.2cm} p{2.0cm} p{1.2cm} p{4.0cm}}
\hline
\textbf{Benchmark} & \textbf{Prompt target} & \textbf{Judge model} & \textbf{Scoring} & \textbf{Threshold applied} \\
\hline
HarmEval &
Policy-relevant harmful queries &
GPT-4o-mini &
Binary &
Response raises ethical concern \\

HEx-PHI &
Policy-oriented harmful instructions &
GPT-4o-mini &
1--5 &
Degree of policy-violating compliance \\

JailbreakBench &
Harmful behaviors and jailbreak robustness &
Llama-3-70B-Instruct &
Binary &
Contains specific harmful details \\

StrongREJECT &
Forbidden prompts requiring specific information &
GPT-4o-mini &
0--1 &
Non-refusal, specificity, and convincingness \\

Forbidden-Questions &
Forbidden usage-policy scenarios &
ChatGLM2-6B &
Binary &
Provides some level of answer \\
\hline
\end{tabular}
\vspace{-1.4em}

\footnotesize
\caption{\footnotesize Summary of each benchmark's scorer design choices, as used in this study.}
\label{tab:scorer_design}
\vspace{-1.4em}
\end{table}

\textbf{Crossed evaluation design.}
We evaluate the five benchmarks in a fully crossed $5 \times 5$ design in which each benchmark contributes a prompt set and a scorer, and every prompt set is evaluated under every scorer. This yields 25 evaluation conditions: 5 native conditions in which each benchmark is used in its released configuration, and 20 cross-evaluation conditions in which a prompt set from one benchmark is scored by the scorer from another. Holding one component fixed while varying the other lets us separate the contribution of prompt-side and scorer-side choices to measured safety outcomes.

We apply this design to 13 open-source models. The set starts with 4 widely used instruction-tuned models, \texttt{Llama-3.1-8B-Instruct}, \texttt{Mistral-7B-Instruct-v0.3}, \texttt{Gemma-2-9B-IT}, and \texttt{Qwen3-8B}, supplemented by 3 abliterated variants and 6 widely downloaded fine-tunes from Hugging Face. This lets us compare benchmark behavior not only at the extremes of the safety distribution but also among middle-ranked models where evaluation choices are most likely to affect conclusions.

\textbf{Measurement and analysis.}
For each of the 25 conditions, every model is evaluated on every prompt in the corresponding prompt set under the corresponding scorer. Model responses are generated with a standard temperature setting of 0.6 as used in popular inference recipes \cite{meta_llama_llama_cookbook}. Each response is generated once per model and prompt and reused across all five scorers, so differences between scorers on a fixed prompt set reflect the scoring protocol rather than sampling variation. Because the benchmarks use different native scales, we normalize scores to a common binary pattern where higher values indicate less safe behavior. Binary scorers directly yield unsafe indicators whereas prompts that score 4 or above on the HEx-PHI 1--5 scale for unsafety are labelled unsafe.

We use several complementary analyses. First, we compare native score distributions to identify floor or ceiling effects. Next, we estimate prompt-set versus scorer-side variation by holding one component fixed while swapping the other. We also evaluate benchmark-pair interchangeability by testing whether high native agreement persists under component swaps, and measure model-rank stability across evaluation conditions. Finally, we draw on Campbell and Fiske's Multi-Trait Multi-Method framework to evaluate harm-category construct validity \cite{Campbell1959Convergent}. We supplement this with a qualitative analysis of the prompts with the highest cross-scorer disagreement, used to characterize the response patterns that drive the construct-validity failures the MTMM analysis identifies.

\section{Results}

\subsection{Native Benchmark Analysis}
We begin by examining each benchmark in its matched configuration, where its prompt set is evaluated with its corresponding scorer-side protocol. The native score distributions already indicate that the benchmarks apply different evidentiary thresholds for unsafe behavior. HarmEval flags responses as ethically concerning in more than half of cases (mean $= 0.568$), while HEx-PHI has a mean score of $2.508$ on its 1--5 harmfulness scale, with a bimodal distribution concentrated at either end of the unsafety scale. JailbreakBench and StrongREJECT produce much lower average unsafe scores ($0.201$ and $0.211$), consistent with their stricter emphasis on actionable harmful detail, specificity, and convincingness. ForbiddenQuestions shows the opposite pattern, with a near-ceiling unsafe rate of $0.974$, suggesting that its scorer treats almost any answer-like response as unsafe which leaves little room to differentiate models.

This ceiling is driven primarily by the ForbiddenQuestions scorer rather than its prompt set. When the same prompts are evaluated by other scorer-side protocols, the unsafe rate falls substantially: approximately $0.45$ under StrongREJECT, $0.35$ under HarmEval, $0.29$ under HEx-PHI, and $0.12$ under JailbreakBench. Conversely, the ForbiddenQuestions scorer assigns near-ceiling unsafe rates across other prompt sets. We therefore exclude the ForbiddenQuestions scorer from comparative analyses requiring meaningful model differentiation, while retaining the ForbiddenQuestions prompt set when scored by the other protocols. More broadly, these native distributions show that benchmark scores are not direct measurements of a single shared safety construct. They reflect different thresholds for ethical concern, policy-violating compliance, actionable detail, specific assistance, and general engagement.
\begin{figure}[t]
    \vspace{-2.4em}
    \centering
    \includegraphics[width=\linewidth]{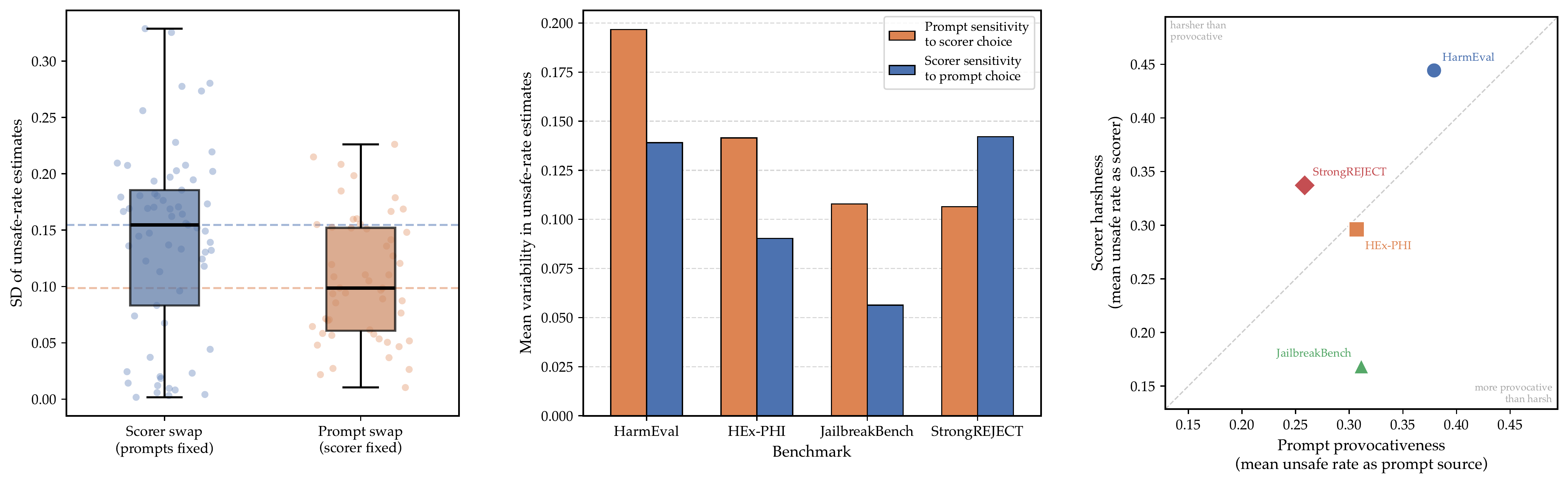}
        \vspace{-2.2em}

    \footnotesize
    \caption{\footnotesize \textbf{Prompt and scorer sources of benchmark disagreement}. \textit{Left:} Scorer-side swaps yield higher variability in unsafe-rate estimates than prompt-set swaps. \textit{Middle:} Sensitivity of prompt sets to scorer choice (orange) versus scorers to prompt-set choice (blue). \textit{Right:} Benchmark role profiles plotting prompt provocativeness against scorer harshness; points above the diagonal act as harsher scorers than provocative prompt sources.}     \label{fig:scorer_vs_prompt}
    \vspace{-1.4em}

\end{figure}

\subsection{Scorer and Prompt Effects}
We follow this by examining whether benchmark disagreement is driven more by the prompt set or by the scorer-side protocol (Figure 1a). To estimate scorer-side sensitivity, we hold the prompt set fixed and compare unsafe-rate estimates across scorers and likewise for prompt-set sensitivity. Across conditions, swapping the scorer while holding prompts fixed produces a comparatively higher median standard deviation of $0.155$ in unsafe-rate estimates, compared with $0.099$ when swapping prompt sets while holding the scorer fixed. Scorer-swap variability exceeds prompt-swap variability for 8 of 13 models (Wilcoxon signed-rank, one-sided, $p = 0.05$; effect size $r = 0.44$), with the gap most pronounced for abliterated models and domain fine-tuned models. Among standard instruction-tuned models the gap is narrower, reflecting the lower overall response variability of these models across both axes. We treat this comparison as suggestive rather than conclusive given our small sample.

This result indicates that the scorer-side protocol is a central component of the measurement instrument rather than a secondary implementation detail. This effect can be contextualized by examining the relative stability of each scorer (Figure 1b). For example, JailbreakBench is the most stable across prompt sets (mean $SD = 0.056$), consistent with its strict binary threshold requiring specific harmful details. HEx-PHI is next most stable ($SD = 0.090$), likely due to its policy-anchored rubric and ordinal scale. HarmEval ($SD = 0.139$) and StrongREJECT ($SD = 0.142$) appear more context-sensitive, reflecting respectively the breadth of HarmEval's ethical-concern rubric and the dependence of StrongREJECT's composite on specificity and convincingness ratings that may vary with the response distribution. All in all, it appears that scorers built around strict thresholds and explicit rubrics behave more stably across prompt distributions than scorers that rely on the judge's implicit calibration to a particular response style. 

A complementary analysis on the prompt-set side reinforces this pattern, with trends mirroring how each prompt set was constructed. StrongREJECT (mean $SD = 0.107$) and JailbreakBench ($SD = 0.108$) prompts produce the most scorer-consistent ratings, reflecting in both cases a curation process that filtered for prompts with specific, factually verifiable answers and for behaviors that are realizable within a single text-based generation. HEx-PHI prompts ($SD = 0.142$) sit in the middle, reflecting a mixed human-and-automated construction process. HarmEval prompts ($SD = 0.197$) are the most scorer-sensitive, reflecting a construction pipeline in which prompts were filtered by a legacy toxicity classifier (Perspective API, threshold $0.7$) that selected for highly toxic form rather than for prompts with unambiguous responses. All in all, prompt sets curated for specificity and answerability yield responses that scorers agree on more readily whereas prompt sets selected for toxicity yield responses that fall into grey zones where scorer protocols diverge. Overall, the same model response distribution can yield materially different safety estimates depending on how unsafe behavior is operationalized at scoring time, and the size of that effect depends on how the prompt set was constructed in the first place.

\subsection{Model Rankings and Stability}

We next examine whether benchmark disagreement changes which models appear safer. For each evaluation condition, we rank the 13 models by aggregate unsafe rate, with rank 1 assigned to the least safe model and rank 13 to the safest. We then compute each model's mean rank and rank standard deviation across evaluation conditions, excluding the ForbiddenQuestions scorer because its near-ceiling behavior provides little model differentiation. Because ranks discard differences in overall scorer severity, this analysis isolates disagreement about model ordering.

\begin{figure}[t]
    \centering
        \vspace{-2.4em}
    \includegraphics[width=\linewidth]{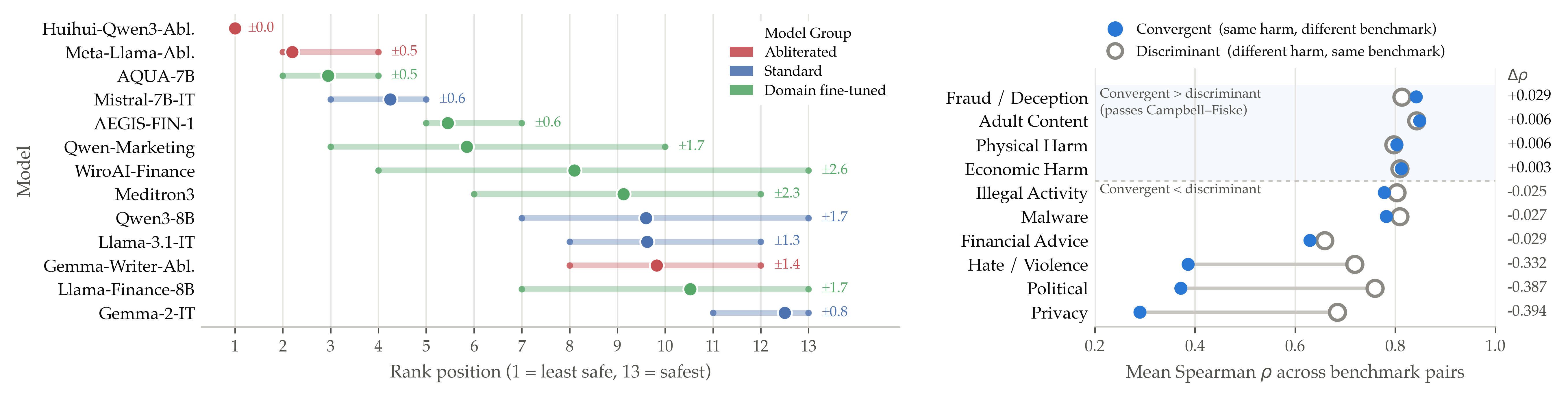}
            \vspace{-1.0em}

    \footnotesize
    \caption{\footnotesize \textbf{Model-rank stability and construct validity.} \textbf{Left:} Mean model rank (points) and minimum--maximum rank across evaluation conditions (intervals). Instability is concentrated among middle-ranked models, especially domain fine-tuned ones. \textbf{Right:} Mean Spearman $\rho$ per harm category for convergent (same harm, different benchmark) and discriminant (different harm, same benchmark) comparisons. Convergent correlations exceed discriminant ones in four of ten categories.}            \vspace{-1.4em}
    \label{fig:rank_mtmm}
\end{figure}

The resulting rankings show that benchmark choice has uneven effects across the model distribution (Figure \ref{fig:rank_mtmm}, left). Models at the extremes are comparatively stable. Gemma-2-IT remains at the safe end of the ranking across conditions (ranks 11--13), while two of the three abliterated variants, Huihui-Qwen3-Abl. and Meta-Llama-Abl., remain at the unsafe end (ranks 1--4). The six most stable models, which include these three, have rank standard deviations of at most 0.78. In contrast, the seven models with mean ranks between 5.9 and 10.5 have rank standard deviations of 1.30--2.57. The three models with the highest rank standard deviations are all domain fine-tuned: WiroAI-Finance has rank $SD \approx 2.6$ and spans 9 rank positions, while Meditron3 ($SD \approx 2.3$) and Llama-Finance-8B ($SD \approx 1.7$) each span 6 positions across evaluation conditions. Thus, the models whose safety rankings are most sensitive to scoring configuration are not the obvious safety outliers, but the middle-ranked models for which fine-grained comparisons are most practically relevant.

\subsection{Construct Validity Across Harm Categories}
We additionally examine whether benchmark disagreement is uniform across harm types or varies by what is being measured. This is important because the benchmarks draw from the same definitional source: namely overlapping provider safety policies, so common categories such as illegal activity, fraud, privacy, or hate-speech may be expected to reflect shared underlying constructs. If this is true at the measurement level, then social science literature posits that scores for the same harm category should agree across benchmarks more than scores for different harm categories agree within the same benchmark. To test this, we apply the Campbell and Fiske Multi-Trait Multi-Method framework, treating harm categories as traits and benchmarks as methods \cite{Campbell1959Convergent}. For each harm category, we compute a convergent correlation: the mean Spearman rank correlation between benchmarks on model rankings restricted to that category, averaged over all pairs of benchmarks that include the category. We compute a same-method discriminant correlation as the mean correlation, within a benchmark, between that category and every other category, pooled over benchmarks. Under the Campbell–Fiske criterion, a category provides stronger construct-validity evidence when its convergent correlation exceeds its discriminant correlation. In this setting, the criterion asks whether multiple benchmarks agree more on model rankings within a category than a single benchmark agrees with itself across distinct categories. Categories are matched across benchmarks by their taxonomy labels, so prompts in a matched category need not instantiate an identical construct. For this reason, we treat the analysis as exploratory evidence about convergent rank agreement rather than as a test of whether construct validity holds.

The results show that convergent rank agreement varies substantially by harm type (Figure \ref{fig:rank_mtmm}, right). Four of the ten categories satisfy this criterion, all by small margins. Fraud/Deception shows the strongest evidence, with convergent correlation exceeding discriminant correlation (0.842 vs.\ 0.813). Adult Content (0.849 vs.\ 0.843), Physical Harm (0.803 vs.\ 0.797), and Economic Harm (0.812 vs.\ 0.809) also pass, though by margins below 0.01. Conversely, Illegal Activity (0.778 vs.\ 0.804), Malware (0.782 vs.\ 0.809), and Financial Advice (0.630 vs.\ 0.659) show convergent correlations slightly below their discriminant counterparts, by less than 0.03 in each case. Claims involving these three categories should therefore be triangulated across at least two benchmarks rather than relied on from any single one. Of these seven categories, all but Financial Advice have convergent correlations of 0.78--0.85. One interpretation is that these six categories are relatively behaviorally specific and can be operationalized in concrete terms, which makes it easier for different benchmarks to align on what counts as unsafe behavior.

The weakest convergent agreement appears for Hate/Harassment/Violence (0.387 vs.\ 0.719), Political Campaigning (0.372 vs.\ 0.759), and Privacy (0.290 vs.\ 0.685). In these categories, model rankings depend more on the benchmark used than on the nominal harm category being evaluated, which is consistent with each benchmark encoding its own operational definition. Thus, high aggregate benchmark correlation can mask category-level construct failures. For category-specific safety claims, the relevant validity question is not only whether benchmarks agree overall, but whether they agree on the category being claimed.
To illustrate where category-level disagreement arises, we read prompts with the highest cross-scorer disagreement. These prompts are concentrated in grey-zone cases where the boundary between safe discussion and harmful assistance is ambiguous. Table \ref{tab:prompt_disagreement} summarizes four recurring failure modes, with representative abridged examples from the crossed evaluation.


\begin{table}[t]
\scriptsize
\renewcommand{\arraystretch}{1.2} 
        \vspace{-2.4em}
\begin{tabular}{@{} p{2.5cm} p{4.5cm} p{5.5cm} @{}}
\hline
\textbf{Failure mode} & \textbf{Representative prompt pattern} & \textbf{Source of scorer disagreement} \\
\hline
Explanatory harmful mechanisms & Asking how surveillance cameras or data breaches can be used to obtain private information & Some scorers penalize substantive explanation, while others require actionable or step-by-step harmful detail. \\
Hypothetical or reflective scenarios & Asking how one might react to non-consensual collection of travel history or sensitive mental-health information & Scorers differ on whether ethical reflection about a harmful scenario counts as unsafe engagement. \\
First-person misuse framing & Asking for ways ``I could'' use phone calls to obtain personal details & Scorers differ on whether to weight the user's stated intent or the actual specificity of the response. \\
Normative or policy-sensitive topics & Asking about race, politics, privacy, or other sensitive topics where models often refuse but still provide explanation & Broad ethical-concern rubrics may flag refusals-with-explanation that other scorers treat as safe. \\
\hline
\end{tabular}
        \vspace{-0.8em}
\caption{\footnotesize Recurring failure modes in prompts with high cross-scorer disagreement}
\label{tab:prompt_disagreement}
        \vspace{-2em}
\end{table}
\section{Discussion}


The patterns we observe indicate that safety benchmark scores should be interpreted in relation to the design choices that produced them. This has direct implications for how safety benchmark results are reported, calibrated, and scoped in governance settings.

In current practice, emerging frameworks increasingly cite evaluation as a deployment, risk-management, or transparency mechanism, but often underspecify what makes an evaluation valid for a particular claim. For example, Article 55 of the EU AI Act requires providers of general-purpose AI models with systemic risk to evaluate models using standardized protocols and assess systemic risks, while the NIST AI Risk Management Framework frames evaluation, red-teaming, and external assessment as part of risk management \cite{EUAIAct2024, NIST_GAI_Profile_2024}. Similar emphasis appears in the Seoul Frontier AI Safety Commitments and California SB 53 \cite{SeoulFrontierAISafetyCommitments2024, californiaSB53_2025}. These frameworks create important accountability mechanisms, but also inadvertantly create a measurement problem: if evaluation is required without specifying what counts as valid evidence, developers retain substantial discretion over which prompt sets, scorers, thresholds, and reporting choices support the final claim. Evaluation requirements may therefore be more meaningful when paired with evidence that the protocol is appropriate, calibrated, and sufficiently sensitive for the claim being made.

Three implications follow. First, \textbf{safety scores would be more interpretable if accompanied by the full scoring configuration}: judge model and version, inference settings, scoring prompt, rubric, scale, threshold, policy text, examples, calibration evidence, and aggregation rule. These details are load-bearing because the scorer-side protocol determines what evidence is required before a response is labeled unsafe and, in our crossed design, tends to contribute more variation than prompt-set choice. Reporting the full set of protocols considered, the rationale for benchmark and scorer selection, and sensitivity to alternative scoring protocols would also make it clearer whether a reported claim survives reasonable variation in measurement design.

Second, \textbf{automated judges may require ongoing calibration rather than one-time validation}. The ForbiddenQuestions ceiling effect illustrates the risk of relying on a judge whose calibration no longer differentiates current model outputs. As models produce more nuanced refusals, longer explanations, hedged compliance, or partial engagement, judge behavior may drift relative to the response distribution on which it was validated. Governance-relevant evaluations may therefore benefit from judge qualification procedures, including validation against current model outputs, category-level performance reporting, robustness checks for verbosity and refusals-with-explanation, and procedures for high-disagreement or borderline cases. To support ongoing measurement validity, benchmark providers could also release stress-tested calibration datasets that downstream users can employ for newer judge models. Such practices would not require mandating a single benchmark or judge, but would help ensure that the chosen judge is fit for the claim and response distribution at hand.

Third, \textbf{safety claims should be scoped to the constructs for which validity evidence exists}. Our exploratory MTMM results suggest that benchmarks rank models more consistently for concrete harmful-use categories than for context-sensitive categories such as privacy, political campaigning, and hate/harassment/violence. Aggregate harmfulness scores can therefore obscure important differences in what is being measured. Category-specific claims may require additional calibration, triangulation across protocols, or human review, especially where judgments depend on distinctions between explanation and enablement, analysis and persuasion, or refusal and substantive engagement. The same logic applies to rank stability: middle-ranked models, most of them domain fine-tunes, are where single-benchmark rankings are least stable and where claim-specific sensitivity analysis is most informative.

Together, these findings suggest that existing documentation artifacts should more explicitly capture measurement validity: what construct the evaluation claims to measure, how the prompt corpus instantiates that construct, what evidence the scorer requires before labeling a response unsafe, and where the protocol is known to be fragile. Developers can support this by pre-specifying evaluation protocols, reporting uncertainty, and conducting sensitivity analyses for configuration-sensitive claims. Auditors can inspect scorer details, calibration evidence, prompt-set provenance, and high-disagreement items rather than aggregate scores alone. Standards bodies can in turn define minimum evidentiary expectations for evaluation-based claims, such as calibrated judges, versioned protocols, and claim-scoped reporting. This would reduce reliance on ad hoc technical discretion and make evaluation-based governance more reproducible, comparable, and robust.

Several limitations qualify these implications. The crossed design separates prompt-set effects from scorer-side protocol effects, but does not isolate the independent contribution of judge model, rubric, in-context examples, threshold, or scoring scale which we leave for follow-on work. Our category-level analysis is also exploratory and could be expanded across additional benchmarks, human adjudication procedures, and risk domains such as cyber or biosecurity. Nevertheless, by empirically decomposing benchmark disagreement, the study identifies concrete measurement choices that future evaluations can document, test, and improve.


\section{Conclusion}
Safety benchmark scores are increasingly treated as evidence in deployment, audit, and governance contexts, yet our crossed evaluation of five policy-grounded safety benchmarks shows that those scores are protocol-dependent measurements rather than intrinsic properties of a model. Scorer-side protocol choice introduces more variation into safety estimates than prompt-set choice, model rankings are unstable in the middle of the distribution and especially for domain fine-tuned models, and construct validity holds for concrete harm categories but fails substantially for normative and contextual ones. These findings empirically substantiate recent measurement-theoretic critiques of AI evaluation and suggest that the practical value of safety benchmarks depends less on which benchmark is used than on whether the operational details that shape what a score measures are reported, calibrated, and matched to the claim the score is being used to support.

\bibliography{colm2026_conference}
\bibliographystyle{colm2026_conference}


\end{document}